\documentclass[a4paper,preprint,superscriptaddress,showpacs,showkeys,nofootinbib,11pt]{revtex4-1}
\usepackage[english]{babel}
\usepackage[dvips]{graphicx}
\usepackage{xcolor,url,amsthm,amssymb,amsmath,mathtools,latexsym}

\begin{document}

\preprint{SB/F/478-18}

\title{Local available quantum correlations for Bell Diagonal states and markovian decoherence}

\author{Hermann L. Albrecht Q.}
\affiliation{Departamento de F\'{\i}sica, Universidad Sim\'on Bol\'{\i}var, Apartado postal 89000, Caracas 1080,
Venezuela.}
\email[Corresponding Author: ]{albrecht@usb.ve}

\author{Douglas F. Mundarain}
\email{dmundarain@ucn.cl}
\affiliation{Departamento de F\'{\i}sica, Universidad Cat\'olica del Norte, Casilla, 1280 Antofagasta, Chile.}

\author{Mario I. Caicedo S.}
\email{mcaicedo@usb.ve}
\affiliation{Departamento de F\'{\i}sica, Universidad Sim\'on Bol\'{\i}var, Apartado postal 89000, Caracas 1080,
Venezuela.}

\begin{abstract}
Local available quantum correlations (LAQCs), as defined by Mundarain et al. \cite{LAQCs}, are analytically determined for Bell Diagonal states. Using the Kraus operators formalism \cite{Kraus}, we analyze the dissipative dynamics of 2-qubit LAQCs under markovian decoherence. This is done for Werner states under the depolarizing \cite{Nakahara} and phase damping channels \cite{Nielsen}. Since Werner states are among those that exhibit the so called \emph{entanglement sudden death} \cite{EntSuddenDeath}, the results are compared with the ones obtained for Quantum Discord \cite{qDiscord}, as analyzed by Werlang et al. \cite{qD-SuddenDeath}, as well as for entanglement, i.e. Concurrence \cite{Wooters9798}. The LAQCs quantifier only vanishes asymptotically, as was shown to be the case for Quantum Discord, in spite of being lower.
\end{abstract}

\keywords{Quantum correlations, Quantum discord, Entanglement, 
Bell Diagonal states, Werner states, Decoherence, Kraus operators.}

\pacs{03.65.Ud, 03.65.Yz, 03.67.Mn.}

\maketitle

\section{\label{sec:Intro}Introduction}

The study of quantum correlations is at the core of Quantum Information Theory (QIT). Entanglement \cite{Horodecki-Ent} had been considered to solely encompass what Schr\"odinger himself esteemed to be ``\emph{the} characteristic trait of quantum mechanics, the one that enforces its entire departure from classical lines of thought'' \cite{ES}. The development of Quantum Discord (QD) by Olliver and Zurek, and independently by Henderson and Vedral \cite{qDiscord}, in 2001 showed that there are quantum correlations that are not included within the separability criteria of entanglement. Using Werner states as an example, both articles show that there are states that are not entangled, i.e. null concurrence \cite{Wooters9798}, and yet exhibit nonzero QD. This has given a new impulse to a highly dynamical subfield of QIT, the study of new quantifiers for quantum correlations.

Local measurements are the key ingredient to properly define correlations. They are important because correlations must quantify the ability of one local observer to infer the results of a second local observer from his own local results.  The aforementioned Quantum Discord \cite{qDiscord}:

\begin{equation}\label{eq:QD-Def}
D_A(\rho_{AB})\equiv \min_{\{\Pi_i^A\}}\left\{I(\rho_{AB})-I[(\Pi^A\otimes\mathbf{1})\rho_{AB}]\right\} = \min_{\Omega_0}\left[I(\rho_{AB})-I(\rho^{cq}_{AB})\right]
\end{equation}

\noindent{}is based on comparing the quantum Mutual Information, defined for the original state $\rho_{AB}$ as

\begin{equation}\label{eq:InfoMutua}
I(\rho_{AB})\equiv S(\rho_A)+S(\rho_B)-S(\rho_{AB})
\end{equation}

\noindent{}with a corresponding classical-quantum (or A-classical) state

\begin{equation}\label{eq:Estados_Cl-Q}
\rho^{cq}_{AB}=\sum_i\, p_i\,|i\rangle\langle i|\otimes\rho_B^i = \sum_i\, p_i\,\Pi^{(A)}_i\otimes\rho_B^i
\end{equation}

\noindent{}which is a postmeasurement state in the absence of readout, where the measurement is performed locally over the A subsystem of $\rho_{AB}$. Analogously, one can define $D_B(\rho_{AB})$ comparing with a quantum-classical (or B-classical) state

\begin{equation}\label{eq:Estados_Q-Cl}
\rho^{qc}_{AB}=\sum_i\, p_i\,\rho_A^i\otimes|i\rangle\langle i| = \sum_i\, p_i\,\rho_A^i\otimes\Pi^{(B)}_i
\end{equation}

\noindent{}Quantifiers of quantum correlations using either A-classical or B-classical states are called Discords and are, in general, not symmetrical.

Other quantifiers \cite{Modi-qDiscord} are based on the difference of a quantity (e.g. mutual information, relative entropy, etc.) with respect to systems in which both subsystems have been locally measured. These type of states are labeled as strictly classical 

\begin{equation}\label{eq:Estados-Clasicos}
\rho_{AB}^c = \sum p_{ij} |\phi_i\rangle_A\langle\phi_i|\otimes |\psi_j\rangle_B\langle\psi_j|
\end{equation}

\noindent{}where $\langle\phi_i|\phi_j\rangle=\delta_{ij}$, $\langle\psi_i|\psi_j\rangle=\delta_{ij}$, $\forall$ $i,j$. It is said that there exists a local basis for which $\rho_{AB}^c$ is diagonal. A special case of strictly classical states \eqref{eq:Estados-Clasicos} worthy of mention are product states, $\rho_{AB}^{\Pi}=\rho_A\otimes\rho_B$. For these type of states, the coefficient $p_{ij}$ in eq. \eqref{eq:Estados-Clasicos} needs to be factorizable, $p_{ij} = p_i p_j$. That is

\begin{equation}\label{eq:Estados_Producto}
\rho_{AB}^{\Pi}=\rho_A\otimes\rho_B = \left[\sum p_{i} |\phi_i\rangle_A\langle\phi_i|\right]\otimes \left[\sum p_{j}|\psi_j\rangle_B\langle\psi_j|\right] = \sum p_{i}p_{j} |\phi_i\rangle_A\langle\phi_i|\otimes |\psi_j\rangle_B\langle\psi_j|
\end{equation}

Quantifiers of this sort include Measurement-Induced Disturbance (MID), introduced by Luo \cite{Luo-MID}, as well as its ameliorated form (AMID), introduced by Wu et al. \cite{AMID-Wu}.

General quantum correlations defined in terms of local bipartite measurements were considered recently by Wu et al. in \cite{Wu}, where they introduce and study non-symmetric quantum correlations using the Holevo quantity \cite{Holevo} and, in a brief final appendix, they define symmetric quantum correlations in terms of mutual information. The LAQCs developed in \cite{LAQCs} focused on a slightly different version of those symmetric correlations, preserving the requirement that any available ones must always be defined in terms of mutual information of local bipartite measurements.

This work is focused on analytically calculating the LAQCs quantifier for the family of Bell Diagonal states, given by

\begin{equation}\label{eq:rho_BD}
\rho^{BD}=\frac{1}{4}\left(\mathbf{1}\otimes\mathbf{1}+\sum c_{i}\sigma_i\otimes\sigma_i\right)
\end{equation}

\noindent{}where the coefficients $c_i\in[-1,1]$ are such that $\rho^{BD}$ is a well behaved density matrix (i. e. has non-negative eigenvalues) and $\sigma_i$ are the well known Pauli matrices, and giving a first glimpse into its dissipative dynamics. This is done by assuming markovian decoherence and using the Kraus operator formalism for two particular quantum channels: depolarization \cite{Nakahara} and phase damping \cite{Nielsen}. We will also make use of the Bloch representation for 2-qubits, given by

\begin{eqnarray}\label{eq:RepresentaHS}
\rho    &=& \frac{1}{4}\left( \mathbf{I}_4 + \vec{x}\cdot\vec{\sigma}\otimes\mathbf{I}_{2} + \mathbf{I}_{2}\otimes\vec{y}\cdot\vec{\sigma}+\mathbf{T}\cdot\vec{\sigma}\otimes\vec{\sigma}\right)\nonumber\\
        &=&\frac{1}{4}\left( \mathbf{I}_4 + \sum_{n=1}^3x_n{\sigma_n}\otimes\mathbf{I}_{2} + \sum_{n=1}^3y_n\mathbf{I}_{2}\otimes\vec{\sigma_n}+\sum_{m,n=1}^3 T_{nm}\sigma_n\otimes\sigma_m\right)
\end{eqnarray}

\noindent{}where $\{\vec{x},\vec{y},\mathbf{T}\}$ are the Bloch parameters given by $x_n=\mathrm{Tr}[\rho(\sigma_n\otimes\mathbf{I}_{2})]$, $y_n=\mathrm{Tr}[\rho(\mathbf{I}_{2}\otimes\sigma_n)]$ and $T_{nm}=\mathrm{Tr}[\rho(\sigma_n\otimes\sigma_m)]$.

The present article is structured as follows: In section \ref{sec:LAQCs-2qubits} we review the main results obtained in \cite{LAQCs} by defining our procedure for calculating the local available quantum correlations quantifier. Section \ref{sec:LAQCs_BD} is dedicated to the explicit calculation of this quantifier for Bell Diagonal states. We start by performing the calculation for a highly symmetrical subset of BD states, namely Werner states. These results are then generalized for the whole BD states family. Section \ref{sec:Decoherence} is devoted to the subject of markovian decoherence. We start by presenting the Kraus operators formalism and proceed to analyze two dissipative quantum channels, namely depolarizing \cite{Nakahara} and phase damping \cite{Nielsen}, acting on the set of Werner states and determining the dissipative dynamics of the LAQCs quantifier by means of our previous result for BD states. Finally, section \ref{sec:Concl} is devoted to the summary.

\section{\label{sec:LAQCs-2qubits}Local available quantum correlations for 2-qubits}

A density operator $\rho$ of a bipartite system $AB$ can always be written in terms of different basis

\begin{equation}\label{eq:rho_Bases}
  \rho=\sum_{klmn}\rho_{kl}^{mn}\,|km\rangle\langle ln|\,=\,\sum_{ijpq}R_{ip}^{jq}\,|B(i,j)\rangle\langle B(p,q)|
\end{equation}

\noindent{}where $k, l, m, n \in \{0,1\}$, $\{|km\rangle\}$ is the well-known computational basis, that is, the basis of eigenvector of $\sigma_z$, which is local, and $\{|B(i,j)\rangle\}$ is another local basis, which is equivalent under local unitary transformations to the former one:

\begin{equation}\label{eq:Base_Bij}
  |B(i,j)\rangle=U^\dagger_a\otimes U^\dagger_b |ij\rangle
\end{equation}

Any such basis for the Hilbert space of qubits can be thought of as a new computational basis, i.e. the basis of eigenvector of $\sigma_{\hat{u}}\equiv\vec{\sigma}\cdot\hat{u}$, where $\vec{\sigma}$ is the vector whose components are the Pauli matrices and $\hat{u}$ is a generic unitary vector. The choosing of such direction can depend on various conditions and / or requirements of the system at hand.

Since strictly classical states are states which are diagonal in some local basis, one can define $X_\rho$ as the strictly classical state \eqref{eq:Estados-Clasicos} induced by a measurement which minimizes the relative entropy

\begin{equation}\label{eq:S(rho||X)}
S(\rho||X_\rho)=\min_{\chi_\rho}S(\rho||\chi_\rho)
\end{equation}

\noindent{}where $\chi_\rho^B$ given by

\begin{equation}\label{eq:Chi_rho^B}
\chi_\rho^B = \sum_{ij}\left[\langle B(i,j)|\rho|B(i,j)\rangle\right]\,|B(i,j)\rangle\langle B(i,j)|
\end{equation}

\noindent{}and $S(\rho||\chi)=-\mathrm{Tr}(\rho\mathrm{log}_2\chi)-S(\rho)$. The minimization of such relative entropy is equivalent to finding the optimal basis $\{|B(i,j)^{opt}\rangle\}$ which will then serve as the new computational basis. Local available quantum correlations are then defined in terms of this optimal computational basis.

Whitout loss of generality, the search for $\{|B(i,j)^{opt}\rangle\}$ can be thought of as the search for the optimal local unitary transformations $U^{op}_a\otimes U^{op}_b$ such that

\begin{equation}\label{eq:rho'}
\rho'= U^{op}_a\otimes U^{op}_b\,\rho\,{U^{op}_a}^\dagger\otimes {U^{op}_b}^\dagger = \sum_{ijpq}\left(R^{op}\right)^{jq}_{ip}|ij\rangle\langle pq|,\,\,\, i,j,p,q\in \{0,1\}
\end{equation}

Therefore, analyzing the criteria for minimization of the aforementioned relative entropy is related to the behavior of the coefficients $\left(R^{op}\right)^{jq}_{ip}$. This is done by defining the most general orthonormal base \eqref{eq:Base_Bij} for each subsystem in terms of the original computational base:

\begin{eqnarray}\label{eq:mu_nu}
\mathrm{A:}\,\,\,|\mu_0\rangle &=& \cos\left(\frac{\theta_A}{2}\right)|0\rangle +\sin\left(\frac{\theta_A}{2}\right)e^{i\phi_A}|1\rangle,\nonumber\\
|\mu_1\rangle &=& -\sin\left(\frac{\theta_A}{2}\right)|0\rangle +\cos\left(\frac{\theta_A}{2}\right)e^{i\phi_A}|1\rangle \nonumber\\
\mathrm{B:}\,\,\, |\nu_0\rangle &=& \cos\left(\frac{\theta_B}{2}\right)|0\rangle +\sin\left(\frac{\theta_B}{2}\right)e^{i\phi_B}|1\rangle,\nonumber\\
|\nu_1\rangle &=& -\sin\left(\frac{\theta_B}{2}\right)|0\rangle +\cos\left(\frac{\theta_B}{2}\right)e^{i\phi_B}|1\rangle
\end{eqnarray}

It is important to keep in mind that this process is equivalent to finding the unitary vectors $\hat{u}_A= (\sin\theta_A \cos\phi_A, \sin\theta_A \sin\phi_A, \cos\theta_A)$ and $\hat{u}_B= (\sin\theta_B \cos\phi_B, \sin\theta_B \sin\phi_B, \cos\theta_B)$ as to define the new $\sigma_{\hat{u}_A}\otimes \sigma_{\hat{u}_B}$ whose eigenvectors define the new computational basis.

In this context, Mundarain et al. define the classical correlations quantifier as

\begin{equation}\label{eq:CorrClasC1}
\mathcal{C}(\rho) = S\left(X_\rho||\Pi_{_{X_\rho}}\right)
\end{equation}

\noindent{}where  $\Pi_{_{X_\rho}}$ is the product state \eqref{eq:Estados_Producto} nearest to $X_\rho$. As shown by Modi et al. \cite{Modi-RelativeEntropy}, the relative entropy of a generic state, e.g. $X_\rho$, and its nearest product state, i.e. $\Pi_{_{X_\rho}}$, is the total mutual information \eqref{eq:InfoMutua} of the generic state. Therefore, the previous definition for the classical correlations quantifier may be rewritten as:

\begin{equation}\label{eq:CorrClasC2}
\mathcal{C}(\rho) = I(X_\rho)
\end{equation}

\noindent{}where $I(X_\rho)$ is the mutual information  of the local bipartite measurement associated with $X_\rho$. Since the mutual information may be written as 

\begin{equation}\label{eq:Info_Mutua-Prob}
I(\rho) = \sum_{i,j} P_{\theta,\phi}(i_A,j_B)\,\log_2\left[\frac{P_{\theta,\phi}(i_A,j_B)}{P_{\theta,\phi}(i_A)P_{\theta,\phi}(j_B)}\right]
\end{equation}

\noindent{}where $P_{\theta,\phi}(i_A,j_B) = \langle \mu_i| \otimes\langle \nu_j| \,\rho\,| \mu_i\rangle \otimes| \nu_j\rangle$ are the probability distributions corresponding to $\rho_{AB}$ and $P_{\theta,\phi}(i_A) = \langle \mu_i| \,\rho_A\,| \mu_i\rangle$, $P_{\theta,\phi}(j_B) = \langle  \nu_j| \,\rho_B\,| \nu_j\rangle$ the ones corresponding to its marginals $\rho_A$ and $\rho_B$, the required minimization of the relative entropy \eqref{eq:S(rho||X)} yields a minima for the classical correlations quantifier defined in \eqref{eq:CorrClasC2}. It is straightforward to see from eq. \eqref{eq:rho'} that $P_{\theta,\phi}(i_A,j_B)$ is directly related to $\left(R^{op}\right)^{jq}_{ip}$ when $\{| \mu_i\rangle \otimes| \nu_j\rangle\}$ is the optimal computational basis.

Once the optimal angles $\theta$ and $\phi$ are found and, therefore, the optimal computational basis is defined, the state is rewritten in terms of this new basis. Since local available quantum correlations are defined in terms of complementary basis, we are interested in determining a new unitary vector $\hat{u}_\perp$, contained in the plane orthogonal to our previous $\hat{u}$. To do so, we define a new unitary vector $\hat{u}_{\Phi_i}$ for each subsystem and define the following basis:

\begin{equation}\label{eq:u0-u1}
  |u_0\rangle(\Phi_n)=\frac{1}{\sqrt{2}}\left(|0\rangle_{opt}+e^{i\Phi_n}|1\rangle_{opt}\right),\,\,\,\,
  |u_1\rangle(\Phi_n)=\frac{1}{\sqrt{2}}\left(|0\rangle_{opt}-e^{i\Phi_n}|1\rangle_{opt}\right)
\end{equation}

\noindent{}where $\{|0\rangle_{opt},|1\rangle_{opt}\}$ is the optimal computational basis and the angles $\Phi_n$ define a direction in the plane perpendicular to $\hat{u}$ for each subsystem, as to define our complementary basis \cite{Wu}. In doing so, we are now able to determine the local available quantum correlations, which are quantified in terms of the maximal mutual information for measurements performed on $\vec{\sigma}\cdot\hat{u}_{\Phi_i}$. That is, we compute the following probability distributions

\begin{equation}\label{eq:P_phi}
  P_\Phi(i_a,j_b,\Phi_a,\Phi_b) = \langle u_i| \otimes\langle u_j| \,\rho\,| u_i\rangle \otimes| u_j\rangle
\end{equation}

\noindent{}and by means of \eqref{eq:Info_Mutua-Prob}, we determine the mutual information $I(\Phi_A,\Phi_B)$, which is then maximized.

\section{LAQCs for Bell Diagonal states}\label{sec:LAQCs_BD}
\subsection{Werner States}

As to better illustrate the calculation of the LAQCs quantifier, we start by determining it for a highly symmetrical subset of Bell Diagonal states \eqref{eq:rho_BD}, namely Werner states, $\rho_w$:

\begin{equation}\label{eq:rhoWerner}
\rho_w=z|\Phi^+\rangle\langle\Phi^+|+\frac{1-z}{4}\,\mathbf{I}_4, \,\,\,\,\,z\in[0,1]
\end{equation}
\noindent{}where $z\in[0,1]$ and $|\Phi^+\rangle=\frac{1}{\sqrt{2}}\left(|0\rangle|0\rangle + |1\rangle|1\rangle\right)$ is a Bell state. Notice that \eqref{eq:rhoWerner} is obtained from \eqref{eq:rho_BD} by setting $c_1=-c_2=c_3=z$. It is well known that for these states, $z<1/3$ implies $\rho_w$ is separable. Nevertheless, as was shown by Olliver $\&$ Zurek and Henderson $\&$ Vedral in \cite{qDiscord}, these states have non-vanishing quantum correlations, i. e. their quantum discord is only null for $z=0$. 

The density matrix for the Werner states, using the standard computational matrix, is written as:

\begin{equation}\label{eq:rhow-Matriz}
\rho_w = \frac{1}{4} \left(\begin{matrix}
1+z	& 0		& 0		& 2z	\\
0	& 1-z	& 0		& 0		\\
0	& 0		& 1-z	& 0		\\
2z	& 0		& 0		& 1+z
\end{matrix}\right)
\end{equation}

By means of \eqref{eq:mu_nu}, the elements $R_{ij}$ \eqref{eq:rho_Bases} for the Werner states are obtained:

\begin{eqnarray}\label{eq:Rij-Werner}
R_{00} &=& \langle\mu_0|\otimes\langle\nu_0|\rho_w |\mu_0\rangle\otimes|\nu_0\rangle \nonumber\\ 
&=& \frac{1}{4}+\left[\cos\left(\frac{\theta_A}{2}\right)\cos\left(\frac{\theta_B}{2}\right)\sin\left(\frac{\theta_A}{2}\right)\sin\left(\frac{\theta_B}{2}\right)\cos\left(\phi_A+\phi_B\right)\right]z\nonumber\\
&& +\left[\cos^2\left(\frac{\theta_A}{2}\right)\cos^2\left(\frac{\theta_B}{2}\right)- \frac{1}{2}\left\{\cos^2\left(\frac{\theta_A}{2}\right)+\cos^2\left(\frac{\theta_B}{2}\right)\right\}+\frac{1}{4}\right]z\nonumber\\
R_{10} &=& \langle\mu_1|\otimes\langle\nu_0|\rho_w |\mu_1\rangle\otimes|\nu_0\rangle \nonumber\\ 
&=& \frac{1}{4}-\left[\cos\left(\frac{\theta_A}{2}\right)\cos\left(\frac{\theta_B}{2}\right)\sin\left(\frac{\theta_A}{2}\right)\sin\left(\frac{\theta_B}{2}\right)\cos\left(\phi_A+\phi_B\right)\right]z\nonumber\\
&& -\left[\cos^2\left(\frac{\theta_A}{2}\right)\cos^2\left(\frac{\theta_B}{2}\right)- \frac{1}{2}\left\{\cos^2\left(\frac{\theta_A}{2}\right)+\cos^2\left(\frac{\theta_B}{2}\right)\right\}+\frac{1}{4}\right]z\nonumber\\
R_{01} &=& \langle\mu_0|\otimes\langle\nu_1|\rho_w |\mu_0\rangle\otimes|\nu_1\rangle = R_{10}\nonumber \\
R_{11} &=& \langle\mu_1|\otimes\langle\nu_1|\rho_w |\mu_1\rangle\otimes|\nu_1\rangle = R_{00}
\end{eqnarray}

Since we are using \eqref{eq:Info_Mutua-Prob} to minimize \eqref{eq:S(rho||X)}, all that is needed are the optimal angles $\{\theta_A, \theta_B,\phi_A, \phi_B\}$. First, we use the symmetry under exchange of subsystems A $\leftrightarrow$ B to simplify our previous expressions using $\theta_1=\theta_2=\theta$ and  $\phi_1=\phi_2=\phi$. Using this, equation \eqref{eq:Rij-Werner} may be written in a more compact form as:

\begin{eqnarray}\label{eq:Rij-Werner_theta}
R_{ij} &=& \frac{1}{4}\left[1-(-1)^{i+j}z\right]-(-1)^{i+j}\sin^2\left(\frac{\theta}{2}\right)\cos^2\left(\frac{\theta}{2}\right)\left[1-\cos(2\phi) \right]z
\end{eqnarray}

\noindent{}where $i,j\in\{0,1\}$. In this expression we have that the first term is $(1\pm z)/4$ separated from the sector with the angular dependence. Therefore, our optimization implies obtaining angles that minimize or even cancel out this term for either $R_{00}=R_{11}$ or $R_{10}=R_{01}$. Analyzing the minimum of \eqref{eq:Rij-Werner_theta}, it is found that this occurs for $\theta=\phi=n\pi$ as well as for $\theta=\phi=n\frac{\pi}{2}$. Due to the high symmetry of Werner states, either of these choices is consistent for obtaining the closest strictly classical state to $\rho_w$ and, moreover, the density matrix for these states \eqref{eq:rhoWerner} is invariant under \eqref{eq:rho'} with either choice of $\theta$ and $\phi$. Therefore, it is consistent to measure our classical correlations in the standard computational basis, that is, for $\theta_1=\theta_2=\phi_1=\phi_2=0$, and $P_{\theta,\phi}(i_A,j_B) = \frac{1}{4}\left(1-(-1)^{i+j}z\right)$ and marginal probabilities $P_{\theta,\phi}(i_A) = P_{\theta,\phi}(i_B) = \frac{1}{2}$. Using these expressions, the classical correlations quantifier \eqref{eq:CorrClasC2} may be written as

\begin{equation}\label{eq:Corr_Clas-Werner}
\mathcal{C}(\rho_w) = \frac{1+z}{2}\log_2(1+z)+\frac{1-z}{2}\log_2(1-z)
\end{equation}

To determine the LAQCs quantifier for the Werner states, we need to define the complementary basis. Since we can consistently measure the classical correlations on the $Z$ direction, the complementary basis used will be eigenstates of $\vec{\sigma}\cdot\hat{u}$, where $\hat{u}$ now lies in the $XY$ plane. The probability distributions $P_\Phi(i_A,j_B,\Phi_A,\Phi_B)$  are then determined from \eqref{eq:P_phi}, where we also make use of the symmetry under exchange of subsystems A $\leftrightarrow$ B so that $\Phi_A=\Phi_B=\Phi$, obtaining:

\begin{eqnarray}\label{eq:P_phi-Werner}
P_\Phi(0_A,0_B,\Phi) &=& \frac{1}{4}[1+z\cos(2\Phi)] = P_\Phi(1_A,1_B,\Phi)\nonumber\\
P_\Phi(1_A,0_B,\Phi) &=& \frac{1}{4}[1-z\cos(2\Phi)] = P_\Phi(0_A,1_B,\Phi)
\end{eqnarray}

\noindent{}where once again we have that $P(0_{A(B)})=P(1_{A(B)})=1/2$ for the marginals $\rho_A$ and $\rho_B$. From these expressions it is again straightforward that the maximum is obtained either for $\Phi=n\pi$, with $n=0,1,2$, or for $\Phi=n\frac{\pi}{2}$, with $n=1,3$. By means of \eqref{eq:Info_Mutua-Prob}, the LAQCs quantifier is then

\begin{equation}\label{eq:LAQCs-Werner}
I(\rho'_w) = \frac{1+z}{2}\log_2(1+z)+\frac{1-z}{2}\log_2(1-z)
\end{equation}

Therefore, we have that for Werner states, there is the same amount of classical correlations as there are locally available quantum correlations.

\subsubsection{Comparing with other quantifiers}

We briefly compare our result \eqref{eq:LAQCs-Werner} for the LAQCs quantifier with other quantum correlations quantifiers, such as quantum discord \cite{qDiscord} and concurrence, a quantifier for entanglement.

It is well known that concurrence\footnote{It is important to notice that we are maintaining the usual notation for Concurrence by using the letter $\mathcal{C}$ and in order to distinguish it from our classical correlations quantifier \eqref{eq:CorrClasC2}, we are using the subscript $w$ to denote the Concurrence for Werner states.}, as introduced by Wootters \cite{Wooters9798}, has a simple expression for Werner states, given by:

\begin{equation}\label{eq:Concurrencia_Werner}
\mathcal{C}_w = \max \left\{0,\frac{3z-1}{2}\right\}
\end{equation}

The expression for quantum discord for Werner states is derived from the analytical one obtained by Luo in \cite{Luo-QD} for the more general case of Bell Diagonal states, given by

\begin{eqnarray}\label{eq:QD-BellD}
D_{BD}&=& \frac{1-c_1-c_2-c_3}{4}\log_2(1-c_1-c_2-c_3)+ \frac{1-c_1+c_2+c_3}{4}\log_2(1-c_1+c_2+c_3)\nonumber \\
& &  +\frac{1+c_1-c_2+c_3}{4}\log_2(1+c_1-c_2+c_3)+ \frac{1+c_1+c_2-c_3}{4}\log_2(1+c_1+c_2-c_3) \nonumber \\
& & - \frac{1-c}{2}\,\log_2\left(\frac{1-c}{2}\right)-\frac{1+c}{2}\,\log_2\left(\frac{1+c}{2}\right)
\end{eqnarray}

\noindent{}Using the fact that $c_1=-c_2=c_3=z$, one can readily obtain the desired expression:

\begin{equation}\label{eq:QD-Werner}
D_w= \frac{1-z}{4}\log_2(1-z)-\frac{1+z}{2}\log_2(1+z)+\frac{1+3z}{4}\log_2(1+3z)
\end{equation}

\begin{figure}[t]
\begin{center}
\includegraphics[scale=0.6]{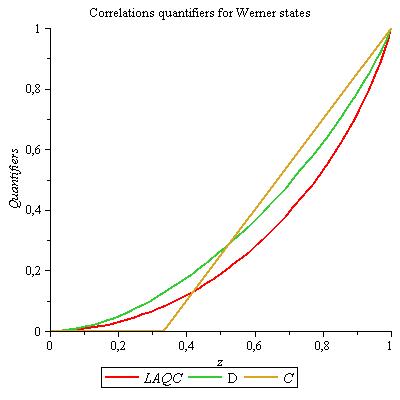}
\caption{\label{fig:Werner-LAQC_QD_C}\small{Quantum correlations quantifiers for the Werner states: LAQC (red line),  Concurrence (yellow line) and Quantum Discord (green line).}}
\end{center}
\end{figure}

Comparison of the LAQCs quantifier with concurrence and quantum discord is shown graphically in Figure \ref{fig:Werner-LAQC_QD_C}. As observed in an example presented in \cite{LAQCs}, the quantifier for the LAQCs has values lower than the ones for Quantum Discord. In the aforementioned case, the 2-qubit pure state $|\psi\rangle=\cos\theta|01\rangle+\sin\theta|10\rangle$, written in the optimal computational basis, exhibits lower values of the LAQCs quantifier for all values of the parameter $\theta$, except for $\theta=0,\pi/2,\pi$, where both quantifiers are null, and for $\theta=\pi/4,3\pi/4$, where both are equal to 1. This same behavior is observed for the Werner states, where both quantifiers exhibit an analogous qualitative behavior, yet the LAQCs quntifier is almost allways lower, except for $z=1$, where both are null, and for $z=1$, where they are maximal, i.e. equal to 1. 

Nevertheless, this does not imply that both quantifiers will necessarily show in general a similar qualitative behavior. As was also pointed out in \cite{LAQCs}, for the family of mixed states $\rho = p|\Psi^-\rangle\langle\Psi^-| +(1-p)|00\rangle\langle 00|$, numerical calculations of both QD and LAQCs quatifiers show, as expected, that the one for LAQCs is less than the one for QD, but also that they behave qualitatively quite differently. Moreover, in the aforementioned work, Mundarain et al. proof that quantum-classical states have null LAQCS, which is not necessarily the case for QD as defined in \eqref{eq:QD-Def}.

\subsection{General case}

We now proceed to the general case of Bell Diagonal states \eqref{eq:rho_BD}. Following the same procedure as before, we determine the coefficients $R_{ij}$:

{\small{\begin{eqnarray}\label{eq:Rij-BD}
R_{00}&=& \langle\mu_0|\otimes\langle\nu_0|\rho_w |\mu_0\rangle\otimes|\nu_0\rangle \nonumber \\ 
&=& \frac{1}{2}\cos\left(\frac{\theta_1}{2}\right)\cos\left(\frac{\theta_2}{2}\right)\sin\left(\frac{\theta_1}{2}\right)\sin\left(\frac{\theta_2}{2}\right) \big[\cos(\phi_1-\phi_2)(c_1+c_2)+\cos(\phi_1+\phi_2)(c_1-c_2)\big] \nonumber\\
& & +\left\{\cos^2\left(\frac{\theta_1}{2}\right)\cos^2\left(\frac{\theta_2}{2}\right)-\frac{1}{2}\left[\cos^2\left(\frac{\theta_1}{2}\right)+\cos^2\left(\frac{\theta_2}{2}\right)\right]+\frac{1}{4}\right\}c_3+\frac{1}{4} \,\,=\,\, R_{11}\nonumber\\
R_{10} &=& \langle\mu_1|\otimes\langle\nu_0|\rho_w |\mu_1\rangle\otimes|\nu_0\rangle \nonumber\\ 
&=& -\frac{1}{2}\cos\left(\frac{\theta_1}{2}\right)\cos\left(\frac{\theta_2}{2}\right)\sin\left(\frac{\theta_1}{2}\right)\sin\left(\frac{\theta_2}{2}\right) \big[\cos(\phi_1-\phi_2)(c_1+c_2)+\cos(\phi_1+\phi_2)(c_1-c_2)\big] \nonumber\\
& & -\left\{\cos^2\left(\frac{\theta_1}{2}\right)\cos^2\left(\frac{\theta_2}{2}\right)-\frac{1}{2}\left[\cos^2\left(\frac{\theta_1}{2}\right)+\cos^2\left(\frac{\theta_2}{2}\right)\right]+\frac{1}{4}\right\}c_3+\frac{1}{4} \,\,=\,\, R_{01}
\end{eqnarray}}}

Since all BD states have maximally mixed marginals, we can again make use of the symmetry under exchange of subsystems A $\leftrightarrow$ B, that is, $\theta_1=\theta_2=\theta$ as well as $\phi_1=\phi_2=\phi$, and rewrite \eqref{eq:Rij-BD} in a more compact form as:
\begin{eqnarray}\label{eq:Rij-BD_theta_1}
R_{ij}= \frac{1}{4}\left[1+(-1)^{i+j} c_3\right]+(-1)^{i+j}\frac{1}{2}\cos^2\left(\frac{\theta}{2}\right)\sin^2\left(\frac{\theta}{2}\right)\left[(c_1+c_2)+\cos(2\phi)(c_1-c_2)-2c_3\right]
\end{eqnarray}

\noindent{}From \eqref{eq:Rij-BD_theta_1} it is straightforward to realize that $\{R_{ii},R_{ij}\}\in[0,1/2]$. 

In this case, the minimization will depend on whether $|c_2|>|c_3|$ or $|c_2|<|c_3|$, that is, on $c_m\equiv\min\{|c_2|,|c_3|\}$. For $c_m=|c_2|$, $\theta=n\frac{\pi}{2}$, with $n=1,2$, and $\phi=\frac{\pi}{2}$, while $\theta=n\pi$, with $n=0,1,2$, and $\phi=n\pi$, with $n=0,1$, for $c_m=|c_3|$. Therefore, we can write our coefficients $R_{ij}^{(opt)}$ as 

\begin{eqnarray}\label{eq:Rij-BD_theta2}
R_{00} = R_{11} = \frac{1}{4}(1+c_m), &\,& R_{10} = R_{01} = \frac{1}{4}(1-c_m)
\end{eqnarray}

As happened for Werner states, due to the symmetry of BD states, the density matrix associated with \eqref{eq:rho_BD} is invariant under the aforementioned unitary transformations \eqref{eq:rho'} for the previously chosen optimal computational basis. Identifying $R_{ij}$ from \eqref{eq:Rij-BD_theta2} as our probabilities distributions $P_{\theta,\phi}(i_A,j_B)$ and the fact that $P(0_{A(B)})=P(1_{A(B)})=\frac{1}{2}$, the classical correlations quantifier \eqref{eq:CorrClasC2} is then given by

\begin{equation}\label{eq:Corr_Clas-BD}
\mathcal{C}(\rho_w) = \frac{1+c_m}{2}\log_2(1+c_m)+\frac{1-c_m}{2}\log_2(1-c_m)
\end{equation}

As previously done for the Werner states, the LAQCs quantifier is then calculated in the basis \eqref{eq:u0-u1}, with $\Phi_A,\Phi_B=\Phi$ due to the symmetry under subsystem exchange A $\leftrightarrow$ B, and the distribution probabilities $P_\Phi(i_A,j_B,\Phi)$ \eqref{eq:P_phi} are then given by:

\begin{eqnarray}\label{eq:P_phi-BD1}
P_\Phi(0_A,0_B,\Phi) &=& \frac{1}{4}\left[1+\frac{c_1+c_2}{2}+\frac{c_1-c_2}{2}\cos(2\Phi)\right] \nonumber\\
P_\phi(1_A,0_B,\Phi) &=& \frac{1}{4}\left[1-\frac{c_1+c_2}{2}+\frac{c_1-c_2}{2}\cos(2\Phi)\right]
\end{eqnarray}

\noindent{}where we also have that $P_\phi(0_A,0_B,\Phi) = P_\phi(1_A,1_B,\Phi)$ and $P_\phi(1_A,0_B,\Phi) = P_\phi(0_A,1_B,\Phi)$. The maximization of \eqref{eq:P_phi-BD1} will now depend on whether $|c_1|>|c_2|$ or $|c_1|<|c_2|$, that is, it will depend on $c_M\equiv\max\{|c_1|,|c_2|\}$. Therefore,

\begin{eqnarray}\label{eq:P_phi-BD2}
c_M=|c_1| \Rightarrow \Phi=n\pi &\Rightarrow& P_\Phi(i_A,j_B)=\frac{1}{4}(1\pm c_1)=\frac{1}{4}(1\pm c_M)\nonumber\\
c_M=|c_2|
\Rightarrow \Phi=n\frac{\pi}{2} &\Rightarrow& P_\Phi(i_A,j_B)= \frac{1}{4}(1\pm c_2)=\frac{1}{4}(1\pm c_M)
\end{eqnarray}

\noindent{}where once again we have that $P(0_{A(B)})=P(1_{A(B)})=1/2$ for the corresponding marginals $\rho_A$ and $\rho_B$. Taking all this into account, the LAQCs quantifier is then

\begin{equation}\label{eq:LAQCs-BD}
I(\rho'_w) = \frac{1+c_M}{2}\log_2(1+c_M)+\frac{1-c_M}{2}\log_2(1-c_M)
\end{equation}

\section{\label{sec:Decoherence}Decoherence}

Modeling the behavior of any real quantum system must take into account that it will not be completely isolated. There will be a much larger system surrounding the quantum one, called environment, which in general will have infinite degrees of freedom. This interaction between quantum system and environment, albeit efforts to minimize it, will induce a process of decoherence and relaxation. This in turn may hinder the ability of the system to maintain quantum correlations, therefore affecting its ability to perform certain tasks in quantum computing, among others. The study of this process can be done, under the markovian approximation, either by using a master equation, i. e. the Lindblad equation \cite{Lindblad}, also referred to as the Lindblad-Kossakowski equation \cite{Kossakowski}, or a quantum dynamical semigroup approach, i.e. Kraus operator \cite{Kraus} formalism. In what follows we will make use of the later, with common interactions to both subsystems, i.e. with the interaction parameter $\gamma$ equal for both subsystems so that:

\begin{equation}\label{eq:Kraus_Interaccion}
  \rho\rightarrow \rho'=\sum_{i,j} \left(\mathbf{E}_i\otimes\mathbf{E}_j\right)\rho\left(\mathbf{E}_i\otimes\mathbf{E}_j\right)^\dagger
\end{equation}

Within this framework, we will study two dissipative quantum channels: Depolarizing \cite{Nakahara} and Phase Damping Channel \cite{Nielsen}.

\subsection{Depolarizing Channel}

This quantum operation represents the process of substituting an initial single qubit state $\rho$ with a maximally mixed one, ${\mathbf{I}}/2$, with probability $1-\gamma$ that the qubit is left unaltered. In terms of the Bloch sphere, the effect of this quantum channel is to uniformly contract the radius of the sphere from 1 to $1-\gamma$ \cite{Nakahara}. Its Kraus operators are given by 

\begin{eqnarray}\label{eq:KrausD}
  \mathbf{E}_0 = \sqrt{1-\frac{3\gamma}{4}}\,\,\mathbf{I}_{2},
    \,\,\,\,\,\,  \mathbf{E}_1 = \frac{\sqrt{\gamma}}{2}\,\,\sigma_x , \,\,\,\,\,\,  \mathbf{E}_2 = \frac{\sqrt{\gamma}}{2}\,\,\sigma_y, \,\,\,\,\,\,  \mathbf{E}_3 = \frac{\sqrt{\gamma}}{2}\,\,\sigma_z
\end{eqnarray}

Applying these operators on a Werner state \eqref{eq:rhoWerner} via \eqref{eq:Kraus_Interaccion}, it is straightforward to verify that the resulting density operator has the following Bloch parameters:

\begin{equation}\label{eq:rhoW_D}
x_n=y_n=0, \forall n;\,\,\,  T_{11} = -T_{22}= T_{33} = z(1-\gamma)^2,\,\,\, T_{mn}=0,\forall m\neq n
\end{equation}

\noindent{}which corresponds to a Werner state where the action of this noisy quantum channel contracts the state parameter $z$ by a factor $(1-\gamma)^2$, that is, it transforms $z\rightarrow z'= z(1-\gamma)^2$. We can now write both classical correlations and LAQCs quantifiers using \eqref{eq:Corr_Clas-Werner} and \eqref{eq:LAQCs-Werner}, obtaining

\begin{eqnarray}\label{eq:LAQC-Werner_Depo}
\mathcal{C}(\rho_w^{Depo}) = I(\rho_w^{Depo}) &=& \frac{1+z(1-\gamma)^2}{2}\log_2[1+z(1-\gamma)^2]\nonumber\\
& &+\frac{1-z(1-\gamma)^2}{2}\log_2[1-z(1-\gamma)^2]
\end{eqnarray}

Let us now compare this with other quantum correlations quantifiers. It is well known that Werner states exhibit entanglement sudden death (ESD) \cite{EntSuddenDeath}, as can easily be seen by using $z'= z(1-\gamma)^2$ in \eqref{eq:Concurrencia_Werner}:

\begin{equation}\label{eq:Concurrencia-Werner_Depo}
\mathcal{C}_w = \max \left\{0,\frac{3z(1-\gamma)^2-1}{2}\right\}
\end{equation}

For quantum discord \cite{qDiscord}, by means of \eqref{eq:QD-Werner} and using $z\rightarrow z(1-\gamma)^2$, the following expression is obtained:

\begin{eqnarray}\label{eq:QD-Werner_D}
  D_w^{(Depo)} &=& \frac{1}{4}\left[1-z(1-\gamma)^2\right]\log_2\left[1-z(1-\gamma)^2\right]-\frac{1}{2}\left[1+z(1-\gamma)^2\right]\log_2\left[1+z(1-\gamma)^2\right]\nonumber\\
  & & + \frac{1}{4}\left[1+3z(1-\gamma)^2\right]\log_2\left[1+3z(1-\gamma)^2\right]
\end{eqnarray}

\begin{figure}[t]
\begin{center}
\includegraphics[scale=0.35]{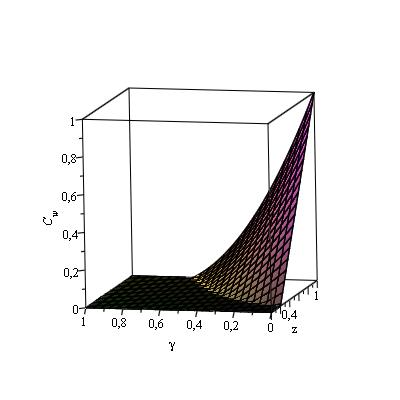} 
\includegraphics[scale=0.35]{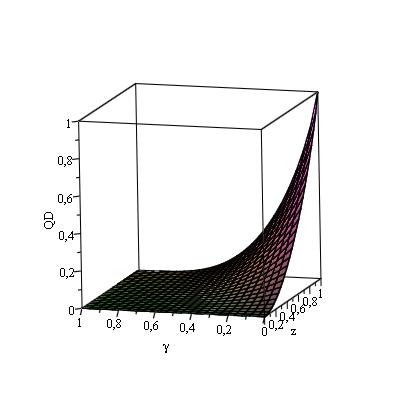} 
\includegraphics[scale=0.35]{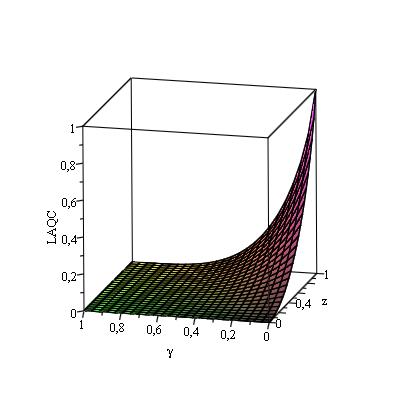}
\caption{\label{fig:Werner_Depo-LAQC_QD_C}\small{Quantum correlations quantifiers for the Werner states under Depolarizing channel: Concurrence, Quantum Discord and LAQC.}}
\end{center}
\end{figure}

The behavior of the LAQCs quantifier, concurrence and quantum discord for a Werner state under the action of a Depolarizing Channel is shown graphically in Figure \ref{fig:Werner_Depo-LAQC_QD_C}. It is worthy noticing that, since the resulting state of this quantum channel is still a Werner state, the qualitative behavior of both QD and LAQCs quantifiers is indeed similar as previously shown, maintaining the relation of the quantifier for QD being greater than the one for LAQCs.

\subsection{Phase Damping Channel}

One of the quantum channels analyzed by Werlang et al. \cite{qD-SuddenDeath} in order to show the robustness of Quantum Disord to decoherence is the \emph{Phase Damping Channel} acting on a Werner state. This noisy channel describes the loss of quantum information without loss of energy \cite{Nielsen}. The Kraus operators for this quantum channel are given by:

\begin{eqnarray}\label{eq:KrausDeph}
  \mathbf{E}_0 = \left(
                    \begin{array}{cc}
                      1 & 0 \\
                      0 & \sqrt{1-\gamma} \\
                    \end{array}
                  \right)
   & \,\, &  \mathbf{E}_1 = \left(
                    \begin{array}{cc}
                      0 & 0 \\
                      0 & \sqrt{\gamma} \\
                    \end{array}
                  \right)
\end{eqnarray}

Applying these operators on a Werner state \eqref{eq:rhoWerner} via \eqref{eq:Kraus_Interaccion}, the resulting density matrix has the following Bloch parameters:

\begin{equation}\label{eq:rhoW_Deph}
   x_n=y_n=0, \forall n;\,\,\, T_{11} = -T_{22} = (1-\gamma)z,\,\,\,  T_{33}=z,\,\,\, T_{mn}=0,\forall m\neq n
\end{equation}

\noindent{}which corresponds to a Bell Diagonal state \eqref{eq:rho_BD} with $c_1=-c_2=(1-\gamma)z$ and $c_3=z$. Since $c_m=\min(|c_{2}|,|c_{3}|)=(1-\gamma)z$ and $c_M=\max(|c_{1}|,|c_{2}|)= (1-\gamma)z$, we can now write our classical correlations and LAQCs quantifiers using \eqref{eq:Corr_Clas-BD} and \eqref{eq:LAQCs-BD}, obtaining

\begin{eqnarray}\label{eq:LAQC-Werner_Deph}
\mathcal{C}(\rho_w^{PD}) =
I(\rho_w^{PD}) &=& \frac{1+(1-\gamma)z}{2}\log_2[1+(1-\gamma)z]\nonumber\\
& & +\frac{1-(1-\gamma)z}{2}\log_2[1-(1-\gamma)z]
\end{eqnarray}

Even though the resulting quantum state is no longer a Werner state, since $c_1 \neq c_3$, we again have an equal distribution of classical and quantum correlations. It is also noticeable that once more there is no 'sudden death' effect observed with the LAQCs quantifier.

Concurrence for \eqref{eq:rhoW_Deph} is given by:

\begin{equation}\label{eq:Concurrence-rho_Deph}
\mathcal{C}_w^{(PD)} = \max\left\{0,\frac{z}{2}\left(3-2\gamma\right)-\frac{1}{2}\right\}
\end{equation}

\noindent{}and Quantum Discord is readily obtained from \eqref{eq:QD-BellD} and \eqref{eq:rhoW_Deph}, yielding:

\begin{eqnarray}\label{eq:QD-Werner_Deph}
  D_w^{(PD)} &=& \frac{1+z(3-2\gamma)}{4}\,\log_2\left[1+z(3-2\gamma)\right]+\frac{1-z(1-2\gamma)}{4}\,\log_2\left[1-z(1-2\gamma)\right]\nonumber\\
 & & -\,\frac{1+z}{2}\,\log_2(1+z)
\end{eqnarray}

\begin{figure}[t]
\begin{center}
\includegraphics[scale=0.35]{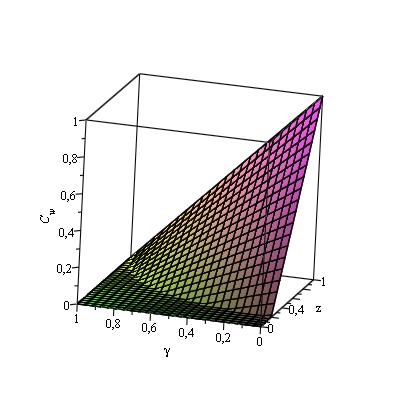} 
\includegraphics[scale=0.35]{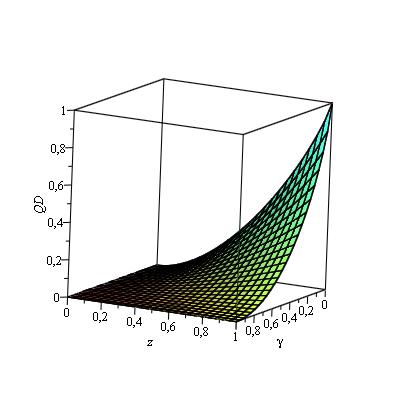} 
\includegraphics[scale=0.35]{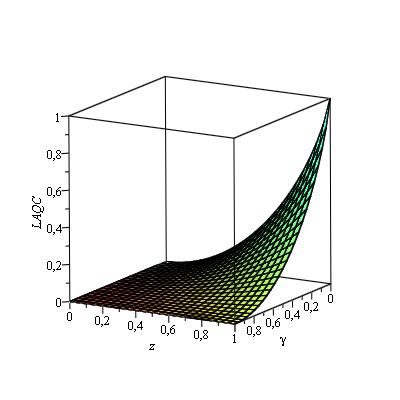}
\caption{\label{fig:Werner_Deph-LAQC_QD_C}\small{Quantum correlations quantifiers for the Werner states under Phase Damping channel: Concurrence, 
Quantum Discord and LAQC.}}
\end{center}
\end{figure}

The behavior of the LAQCs quantifier, quantum discord and Concurrence for a Werner state under the action of a Phase Damping Channel is shown graphically in Figure \ref{fig:Werner_Deph-LAQC_QD_C}. As can be inferred from this graphics, the qualitative behavior of both QD and LAQCs is in this case also quite similar, maintaining the expected relation of QD being larger than LAQCs.

\section{\label{sec:Concl}Conclusions}

We have successfully evaluated the LAQCs quantifier for the family of Bell Diagonal states, obtaining analytical formulas for it. To do so, we started with a much simpler case, the subfamily of Werner states, as to better illustrate the procedure for determining the LAQCs quantifier. For this subset of BD states, its behavior has been graphically presented, comparing it with both concurrence \cite{Wooters9798} and quantum discord \cite{qDiscord}, \cite{Luo-QD}. In this case QD and LAQCs exhibit similar qualitative behavior and, as expected, the LAQCs quantifier is lower in value than QD.

The dissipative dynamics of the 2-qubit LAQCs quantifier under markovian decoherence was studied for Werner states using the Kraus operators formalism in two cases: Depolarizing channel \cite{Nakahara} and Phase Damping channel \cite{Nielsen}. Analytical expressions were obtained for both cases and presented graphically. As  was previously reported for Quantum Discord \cite{qD-SuddenDeath}, LAQCs also do not exhibit the sudden-death behavior shown by entanglement, i.e. concurrence.

\end{document}